# A new analysis of aftershock sequence statistics and fault geometry fingerprinting

Pathikrit Bhattacharya *[1], Bikas K. Chakrabarti [2], Kamal [1]

___________________________________________________________________

[1]Department of Earth Sciences, Indian Institute of Technology, Roorkee-247 667, Uttarakhand, INDIA

[2]Theoretical Condensed Matter Research Division and Centre for Applied Mathematics and Computational Science, Saha Institute of Nuclear Physics, 1/AF, Bidhannagar, Kolkata-700 064, INDIA

E-mail addresses: pathipes@iitr.ernet.in (Pathikrit Bhattacharya), bikask.chakrabarti@saha.ac.in (Bikas. K. Chakrabarti), kamalfes@iitr.ernet.in (Kamal).

*To whom correspondence should be addressed

**We analysed several aftershock sequences around the globe and calculated the cumulative integral of magnitude over time. This quantity is found to have a linear relationship with time having a slope characteristic of the causative fault zone.**

Two well established laws in seismology, dealing with earthquake statistics on global as well as regional scale, are the Gutenberg-Richter law (1) and the Omori Law (2). The Gutenberg-Richter law gives us the cumulative frequency distribution of earthquakes of a given magnitude *M* or above. The Omori law gives us a temporal distribution of aftershocks without reference to the temporal distribution of magnitudes. In this work we present an empirical finding describing the statistics of an aftershock sequence of a specific mainshock. This law therefore describes a regional statistical feature of earthquakes. In particular, we show that the cumulative integral of magnitudes of aftershocks with respect to time elapsed since the occurrence of the mainshock is a straight line whose slope is characteristic of the geometry of the causative fault zone. This empirical law is expected to provide a kind of fractal "fingerprinting" of the fault zone.

We first collected the aftershock magnitude-time sequence $M(t)$ of six events from different catalogs. We then evaluated a cumulative integral $Q(t)$ of the aftershock magnitudes over time. Numerically, we evaluated the integral $Q(t) = \int_0^t M(t')dt'$. A trapezoidal rule was used to evaluate $Q(t)$; here $t$ denotes the time since the main shock. Aftershocks of a major event were considered to be events within a given region, geographically defined as boxes or polygons constrained by suitable latitudes and longitudes, and the magnitudes were recorded over a length of time over which the region has not yet relaxed to its background seismicity.

We acquired several earthquake catalogs from around the globe. In case a catalog was not homogeneous (a homogeneous catalog is one in which all the magnitudes are reported in the same magnitude scale), we used the original data set as well a homogenized data set in which all magnitudes were converted to $M_w$ (moment magnitude) using relations given in (3). These are however global relations and hence might not strictly apply to regional studies but they do serve our primary purpose of establishing the trend. The point will be further discussed when we present our results.

We selected the following events across the globe to emphasize the global validity of our findings:

a. The 1989 Loma Prieta earthquake (18/10/1989, $M_w$ = 7.1, 37.0°, -121.88°). The data set used was the same as the one used in (4).

b. The 1995 Kobe earthquake (17/01/1995, $M_{JMA}$ = 7.2, 34.6°,135.0°). The aftershock region was chosen on the basis of the work done in (5) (latitudes 34°-36°, longitudes 133.5°-137°). The data was taken from the JUNEC catalog for the period 17/01/1995-31/12/1995.

c. The 2004 Parkfield earthquake (28/09/2004, $M_w$ = 5.96, 35.8182°, -120.3660°). The aftershock region was chosen on the basis of suggestions given by the NCEDC (available at www.ncedc.org/2004parkfield.html). The box is defined by latitudes 35.76°-36.06° and longitudes -120.67°- (-120.25°). The data was taken from the NCEDC catalog for the period 28/09/2004-28/02/2008.

d. The 2004 Sumatra earthquake (26/12/2004, $M_w$ = 9.0, 3.30°, 95.98°). The box was chosen in accordance to the earthquake summary poster prepared by the USGS (available at http://earthquake.usgs.gov/eqcenter/eqarchives/poster). The box chosen was latitudes 0°- 20°, longitudes 90°-100°. The data was taken from the USGS (PDE) catalog for the period 26/12/2004-28/02/2008.

e. The Muzaffarabad (Kashmir, North India) earthquake of 2005 (08/10/2005, $M_s$ = 7.7, 34.52°, 73.58°). The box chosen is defined by latitudes 33.5°-35.5° and longitudes 72.2°-74.2°. The data was once again taken from the USGS catalog for the period 08/10/2005-28/02/2008.

f. The 2007 Sumatra earthquake (12/09/2007, $M_s$ = 8.5, -4.44°, 101.37°). The aftershock zone was chosen on the basis of the USGS summary poster (http://earthquake.usgs.gov/eqcenter/eqarchives/poster) for the event and is defined geographically by latitudes 2°- (-8°) and longitudes 96°-108°. This data set was taken from the USGS catalog for the period 12/09/2007-28/02/2008).

All dates are expressed as dd/mm/yyyy, positive latitude implies northern hemisphere, positive longitude implies eastern hemisphere, negative latitude implies southern hemisphere and negative longitude implies western hemisphere. $Q(t)$ for each of the above aftershock sequences were estimated using the above data sets.

| Event No. | $S_1$ | $S_2$ | Power | $S_{loc.1}$ | $S_{loc.2}$ |
|---|---|---|---|---|---|
| **a** | 1.09 | - | 0.98 | 1.22 | - |
| **b** | 2.40 | - | 0.99 | 2.31 | - |
| **c** | 2.44 | 2.43 | 0.98 | 2.31 | 2.93 |
| **d** | 4.47 | 4.83 | 0.99 | 4.59 | 4.92 |
| **e** | 3.91 | 4.45 | 0.99 | 3.95 | 4.44 |
| **f** | 4.63 | 4.97 | 0.99 | 4.45 | 4.78 |

*Table 1. The event tags correspond to those in the text. $S_1$ Corresponds to the slope of the linear fit with the raw data while $S_2$ corresponds to the linear fit with the homogenized data set. The additional subscript loc. for the last two columns give the averages of the local slopes for the raw and homogenized data respectively. For Parkfield, we have assumed the duration magnitude ($M_D$) and the local magnitude ($M_L$) to be equivalent for the given magnitude and depth range. The Loma Prieta and the Kobe data sets were homogeneous.*

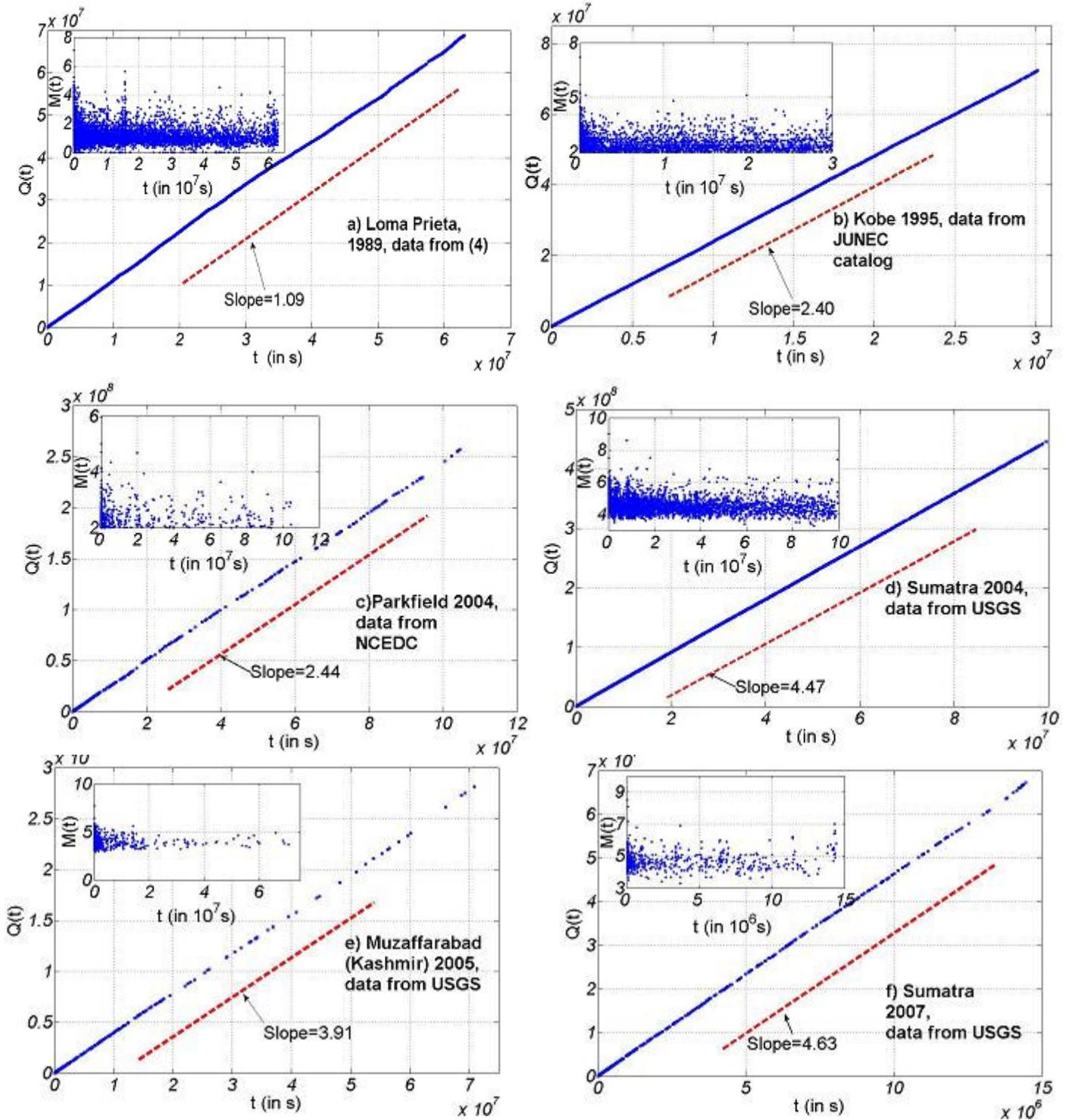

**Fig.1.** Plots of $Q(t) = \int_0^t M(t')dt'$ vs. t for the aftershock sequences of: a) The Loma Prieta quake, 1989, b) The Kobe quake, 1995, c) The Parkfield quake, 2004, d) The Sumatra quake, 2004, e) The Muzaffarabad quake, 2005, f) The Sumatra quake, 2007. The plots are for the raw data. The dashed lines are the straight lines having the indicated slopes and are aids to the eye. The corresponding magnitude-time M(t) series are shown in the respective insets.

Our analyses indicate a clear linear relationship $Q(t) = St$ where $S$ is the slope. Fits are very good in spite of the fact that some of the data sets (Sumatra 2004, Sumatra 2007 and Muzaffarabad 2005 data sets) were heterogeneous with respect to the reported magnitude scales. The Parkfield data set was predominantly homogeneous except a few events. In such cases, of course, the aforementioned global conversion relations were applied and the linear fits were re-estimated. We also tried a power law fit for the $Q(t)$ vs. $t$ statistics and found that the power comes practically unity (Table 1). To further check the fluctuations in the slopes $S$, we calculated local slopes everywhere in the $Q(t)$ time series by taking overlapping segments of 20 points each. This was done for both the raw as well as the processed data sets. The results are presented in Table 1.

The results of our analysis mentioned in Table 1 and the plots in Fig. 1 point clearly to the equation, $Q(t) = St$, $S$ being the slope of the line. The line retains this slope for years. This indicates that $S$ is characteristic of the fault zone. This was further checked by integrating from anywhere in the time series (i.e. shifting our $t = 0$ to any randomly chosen aftershock) after the mainshock. This was done for both the raw as well as the homogenized data. The slopes for such plots were found to be within 2% variability, with respect to the integral evaluated since the mainshock, for all the data sets analyzed. A very interesting case at hand was the unusually large aftershock (date 12/09/2007, 23:49:03.72 UTC, latitude -2.62, longitude 100.84, $M_s = 8.10$) of 12$^{th}$ September for the 2007 Sumatra earthquake (mainshock details: date 12/09/2007, 11:10:26.83 UTC, latitude -4.44º, longitude 101.37º, $M_s = 8.50$). When we integrated onwards from it we found the slopes $S_1 = 4.63$ and $S_2 = 4.96$. That is, when we considered this event as the mainshock, the slopes we obtained (for the raw and the homogenized data sets respectively) were the same (variations are well within the error bounds discussed later). This is very good support for the dependence of the slope on the fault zone and not the magnitude of an individual shock. Also, the 2% variability in slope is clearly within the error bounds induced by the data sets. A wide variety of events can lead to systematic errors in the reported magnitudes; e.g. deviations in the instrumental calibration, changes in the seismic equipment, changes of the agency operating the earthquake recording network, introduction of new software for the analysis, removal or addition of seismograph stations as well as changes in the magnitude definition. Such systematic errors can be very large, as much as 0.5 magnitude units (6). Such errors would set the eventual error bound for the slope as the errors due to fitting are much smaller as mentioned already. With the available catalogs, the errors in slope estimation would be thus 5-7% at least (6, 7).

The robustness of the linear relationship $Q(t) = St$ was further checked by trying to fit the $Q(t)$ data to a power law like $Q(t) = S't^{\gamma}$. We found $\gamma = 1 \pm 0.02$ and $S' \simeq S$. Since the error bar here is much less than the error bounds for the local slopes discussed above, the linear relationship got confirmed. Further, we considered the data set with higher value of lower cut-off of the magnitude from the records (even in some cases by unity), and still obtained the same linear law with unchanged slope (within error limits mentioned). By increasing the lower cut-off, one considers less number of events and also increases the time intervals between the successive events in the series. We find that they cancel in such a way that $Q$ maintains the same linear relationship with unchanged value of the slope.

A physical explanation of the linearity can be obtained from a model described in (8, 9, 10, 11) where the seismic time series is modeled by the time series of overlaps between two fractal sets as one moves with uniform velocity over the other. For some simple regular fractals, this time series can be exactly calculated (10) and the slope $S$ could then be expressed as a function of two parameters: (a) the generation number and (b) the dimensionality of the fractal involved. A particular case can be considered when the overlap sets are obtained out of two identical regular Cantor sets (in relative uniform motion) formed of *(a – 1)* blocks, taking away the central block, giving fractal dimension of each of the Cantor sets equal to *ln(a – 1)/lna*. The $Q(t)$ statistic for this model approaches a strict straight line in the limit $a \to \infty$, asymptotically, with a slope $S$ given by $S = [(a-1)^2/a]^n$ for sets of generation $n$. A good estimation of the generation number $n$ can be obtained from the roughness of the fault geometry (12). Hence a change in slope $S$ would correspond to a change in (a) or (b) or both. This would explain why the slope is such a robust characteristic of the fault zone.

Given an earthquake time series, the slope of the $Q(t)$ vs. $t$ plot thus provides a "fractal fingerprint" of the fault zone responsible for the events. In this regard, however, a point needs to be made about the two Sumatra quakes. The two slopes are indistinguishable given our error bounds though their epicenters are separated by approximately 225 km. This may be due to the extensive fracturing of the region by the 2004 Sumatra quake thus leading to similar fractal characteristics of the two zones.

We have shown, in this paper, that cumulative integral of magnitudes $Q(t)$ of an aftershock sequence over time $t$ is linear, the slope $S$ being characteristic of the fault zone. In the fractal overlap model (8, 9, 10, 11) such linearity is indeed observed and there the slope is a function of the fractal dimensions and the generation numbers of the fractals involved at the fault. Some key features of the fault geometry may therefore be extracted out of such an analysis of the magnitude time series *M(t)* of the earthquake aftershocks.

**References:**

1) Gutenberg, B. & Richter, C. F. Frequency of earthquakes in California. *Bull. Seismol. Soc. Am.* **34**, 185-188 (1944)

2) Omori, F. On the aftershocks of earthquakes. *J. Coll. Imp. Univ. Tokyo* **7,** 111-200, (with Plates IV-XIX) (1894)

3) Scordilis, E. M. Empirical global relations converting $M_s$ and $m_b$ to moment magnitude. *J. Seismol.*, **10,** 225-236 (2006)

4) Kamal & Mansinha, L. The triggering of aftershocks by the free oscillations of the earth. *Bull. Seismol. Soc. Am.* **86** (2)**,** 299-305 (1996)

5) Fig.8 in Toda, S., Stein, R. S., Reasenberg, P. A., Dieterich, J. H. & Yoshida, A. Stress transferred by the $M_w$ = 6.9 Kobe, Japan, shock: Effect on aftershocks and future earthquake probabilities. *J. Geophys. Res.* **103** (B10)**,** 24543-24556 (1998).

6) Castellaro, S., Mulargia, F. & Kagan, Y. Y. Regression problems for magnitudes. *Geophys. J. Int.,* **165** (3), 913-930 (2006)

7) Earthquake Bulletins and Catalogs at the USGS National Earthquake Information Center, Sipkin, S. A., Person, W. J. & Presgrave, B. W. U.S.Geological Survey National Earthquake Information Center, May 2000 (available at http://earthquake.usgs.gov/regional/neic/neic_bulletins.php).

8) Chakrabarti, B. K. & Stinchcombe, R. B. Stick-slip statistics for two fractal surfaces: a model for earthquakes. *Physica A* **270** (1-2)**,** 27-34 (1999)

9) Bhattacharya, P. & Chakrabarti, B. K. (Eds.), *Modelling Critical and Catastrophic Phenomena in Geoscience: A Statistical Physics Approach* (Springer-Verlag, Berlin Heidelberg, 2006)

10) Bhattacharya, P. Geometric models of earthquakes. 155-168 in (9).

11) Pradhan, S. & Chakrabarti, B. K. Search for precursors in some models of catastrophic failures. 459-477 in (9)

12) Hansen, A. & Mathiesen, J. Survey of scaling surfaces. 93-110 in (9)